\magnification=\magstep1
\input amstex
\documentstyle{amsppt}
\voffset-3pc
\baselineskip=24true pt

\topmatter
\title Rigorous Real-Time Feynman Path Integral
       for Vector Potentials
       \endtitle
\author Ken Loo \endauthor
\address PO Box 9160, Portland, OR.   97207
         \endaddress
\email look\@sdf.lonestar.org \endemail
\thanks The author would like to give thanks to Photine Tsoukalas,
        Ray Cote, Xpi, the Pratts, and the referees for their constructive 
        comments.
        \endthanks
\abstract In this paper,
we will show the existence and uniqueness of a  
real-time, time-sliced Feynman path integral for quantum systems 
with vector potential.  Our formulation of the path
integral will be derived on the $L^2$ transition probability
amplitude via 
improper Riemann integrals.  Our formulation will hold
for vector potential Hamiltonian for which its potential
and vector potential each carries at most a finite 
number of singularities and discontinuities.  
\endabstract
\endtopmatter

\document

\define\fac#1#2{\left(\frac {m}{2\pi{}i\hbar\epsilon}\right)^
    {\frac{#1}2\left(#2 + 1 \right)}}

\define\summ#1#2#3#4{\frac {i\epsilon}{\hbar}\sum\limits_{j=1}
    ^{{#1}+1}\left[\frac {m}{2}\left(\frac {{#2} - {#3}}
    {\epsilon}\right)^2\! #4\right]}

\define\pd#1#2{\dfrac{\partial#1}{\partial#2}}
\define\spd#1#2{\tfrac{\partial#1}{\partial#2}}
\define\ham1#1#2{\dfrac{-\hbar^2}{2m}\Delta_{#1} #2}
\TagsOnRight

\subhead
{\bf 1. Introduction}\endsubhead
In his paper(see [22] footnote 13), Feynman observed that
by using wave functions, ill-defined oscillatory integrals
can be given rigorous meaning.  
We will use Feynman's observation and use wave functions to provide
a convergence factor in the derivation of a real time propagator that
takes the form of an $L^2$ transition probability amplitude and the derivation
of a real time, time sliced Feynman path integral.

Since Feynman's invention of the path integral
in the 40's, giving the real-time Feynman path integral
rigorous mathematical justification for general potentials
has been a stumbling block(see [1]-[3], [7]-[8],
[19], and references within).  In physics, the real-time,
time-sliced Feynman path integral is formulated on
the propagator with improper Riemann integrals in hope
of convergence since the integrand of the path integral
in real time is not absolutely integrable in Lebesgue sense
(see [13], [18] and references within).
In the spirit of physics, in a previous work [15], we
formulated a rigorous real-time, time-sliced Feynman path
integral on the $L^2$ transition probability amplitude via
improper Riemann integrals.  Our previous formulation held
for nonvector potential Hamiltonians with potential that 
has at most a finite number of singularities and discontinuities.

In this paper, we will extend our previous work to vector potential
Hamiltonians with potential and vector potential that
carries at most a finite number of singularities and
discontinuities.

In physics, the vector potential Feynman path integral 
for the propagator is formally given by
$$\align
  {}&\lim_{k\to\infty}\left(\dfrac{m}{2i\pi\hbar\epsilon}\right)^
    \frac{n(k + 1)}{2}\int\limits_{\Bbb R^{nk}} 
     \text{exp}\Bigg\{\dfrac{i\epsilon}{\hbar}
      \sum_{j = 0}^k\left[\dfrac{m}{2}
     \left(\dfrac{\vec x_{j+1} - \vec x_j}{\epsilon}\right)^2 -
     V\left(\vec x_j\right)
     \right] + \tag1.1\\
  {}&\dfrac{ie}{\hbar c}\sum_{j = 0}^k
     \left(\vec x_{j+1} - \vec x_j\right)*\vec a\left(
     \frac{\vec x_{j + 1} + \vec x_j}{2}\right)\Bigg\}
     d\vec x_1\dots d\vec x_k ,
\endalign
$$
where the integrals in (1.1) are improper Riemann integrals
(see [13], [18] and references within) and the * inside the 
integral is the vector dot product.  
In this paper, we will derive the following,
$$\align
   {}&\Bigg<\phi^{*},
      \text{exp}\left(-it\bar H\right)
       \psi\Bigg>_{L^2} =
    \int_{\Bbb R^n} \phi\left(\vec x\right)\left[
    \text{exp}\left(-it\bar H\right)
    \psi\right]\left(\vec x\right)
    d\vec x = \tag1.2\\
   {}&\lim_{k\to\infty}\Bigg\{\left(\dfrac{1}{4i\pi\epsilon}\right)^
    \frac{n\left(k - 1\right)}{2}
    \int\limits_{r\Bbb R^{n(k + 1)}}\phi\left(\vec x_k\right)
    \times\\
   {}&\text{exp}\Bigg\{i\epsilon
    \sum_{j = 0}^{k - 1}\left[\dfrac{1}{4}
     \left(\dfrac{\vec x_{j+1} - \vec x_j}{\epsilon}\right)^2 -
     V\left(\vec x_{j+1}\right) +
     \dfrac{\bar \lambda\left(\vec x_{j+1}, \vec x_j\right)}
     {\epsilon}
     \right]\Bigg\}\psi\left(\vec x_0\right)
     d\vec x_0\dots d\vec x_{k}\Bigg\}.
\endalign
$$
In (1.2), we have used 
the notation $r\Bbb R^n$ in the second line
to mean improper Riemann 
integral over $\Bbb R^n$; $\epsilon = \dfrac{t}{k}$;
$\bar H$ is the closure of the essentially self-adjoint
Hamiltonian
$$
  H = \sum_{j = 1}^n\left(-i\nabla_j - \vec a_j\right)^2 + V; \tag1.3
$$
$\phi ,\psi\in L^2$; 
$\phi ,\psi , \vec a , V$ are such that they each have
at most a finite number of discontinuities and singularities, 
and $\bar \lambda$ is dependent on
$\vec a$.

We point out that the function 
$\bar\lambda$
does not
reproduce the term $\left(\vec x_{j+1} - \vec x_j\right)*\vec a\left(
     \dfrac{\vec x_{j + 1} + \vec x_j}{2}\right)$
in (1.1), but we will see that when the vector potential is well behaved
and when $\vec x_{j + 1}, \vec x_j$ are close to each other,
$\bar \lambda\left(\vec x_{j + 1}, \vec x_j\right)$
is close to $\left(\vec x_{j+1} - \vec x_j\right)*\vec a\left(
     \dfrac{\vec x_{j + 1} + \vec x_j}{2}\right)$.
Lastly, both the vector potential and scaler potential
can carry any kind of singularity
as long as the Hamiltonian in 1.3 is essentially self-adjoint.
One possible and interesting application could be the coulomb potential
problem. Duru and Kleinert(see [13] and references within)
formally computed the hydrogen atom propagator by using  
integration over path space version of the
Feynman path integral, they produced the correct energy
eigenvalues.  Kleinert[13] used the time-sliced version
of the Feynman path integral to formally justify the integration
over path space calculation but mathematical rigor is still lacking.

The idea for the derivation of (1.2) is the following.
For suitable operators $A$ and $B$,
Trotter's product formula reads(see [5], [16], and [17])
$$
  \text{exp}\left(-itC\right) =
  s-\lim_{k\to\infty}\left(\text{exp}\left(\dfrac{-itA}{k}\right)
    \text{exp}\left(\dfrac{-itB}{k}\right)\right)^k, \tag1.4
$$
where $C$ is the closure of $A + B$.
Together with the work of Simon (see [19] and lemma 2.4 below), 
it would be reasonable to believe that for the Hamiltonian
in (1.3),
$$
  \text{exp}\left(-it\bar H\right) = 
   \text{s -}\lim_{k\to\infty}\left(
   \text{exp}\left(\dfrac{-itV}{k}\right)
   \prod_{j = 1}^{n}
   e^{i\lambda_j}
   \text{exp}\left(\dfrac{-itH_0}{k}
   \right)e^{-i\lambda_j}\right). \tag1.5
$$
We will use a generalized
Trotter's formula due to Chernoff(see [8]) to prove (1.5)
for (essentially self-adjoint)$H$ 
that satisfies theorem 2.2 or theorem 2.3 below.  From
here on, we will assume that $H$ is such that 
theorem 2.2 or theorem 2.3 below holds.  

We will derive (1.2) with the help of (1.5) and the
following idea.  For simplicity,
suppose $f\left(x\right)\in L^2\left(\Bbb R\right),
g_t\left(x, y\right)
\in L^2\left(\Bbb R\times\Bbb R\right)$ are such that
they are bounded and continuous.  Further, suppose that
both
$$\align
  {}&h_t\left(x\right) = \int_{-a}^{b}
   g_t\left(x, y\right)dy,  \tag1.6\\
  {}&p_t\left(x\right) =
   \text{s -}\lim_{a,b\to\infty}\int_{-a}^{b}
   g_t\left(x, y\right)dy
\endalign
$$
are in $L^2\left(\Bbb R\right)$ as a function of
$x$.  In (1.6), we take the integral to
be Lebesgue integrals and the limits are taken
independent of each other.
Notice that for $p_t\left(x\right)$, we can interpret the
integral
as an improper Lebesgue integral with convergence
in the $L^2$ topology.  Let us denote $\chi_{[-c,d]}$
to be the characteristic function on $[-c, d]$.
Schwarz's inequality then implies
$$\align
  {}&\Bigg|\int_{\Bbb R} f\left(x\right)p_t\left(x\right)dx -
  \int_{-c}^{d}
  \int_{-a}^{b}
  f\left(x\right)g_t\left(x, y\right)dx dy\Bigg| \leq \tag1.7\\
  {}&||f||_2\,||p_t - h_t||_2 + ||f - \chi_{[-c,d]}f||_2\,||h_t||_2
    \to 0.
\endalign
$$
Thus, we can write
$$
  \int_{\Bbb R} f\left(x\right)p_t\left(x\right) =
  \lim_{a,b,c,d\to\infty}\int_{-c}^{d}
  \int_{-a}^{b}f\left(x\right)g_t\left(x, y\right)dx dy, \tag1.8
$$
where the limits are all taken independent of each other.
Since $f$ and $g$ are bounded and continuous,
the Lebesgue integral over $[-a, b]\times [-c, d]$
in (1.8) can be replaced by
a Riemann integral.  Since the limits are taken independent
of each other, we can then interpret the right hand-side
of (1.8) as an improper Riemann integral.  If $f$ and
$g$ carry singularities and discontinuities,
care must be taken in the
region of integration so that
the replacement of Lebesgue integral with Riemann integrals
can be done.  What we have demonstrated 
 in our simple example
is just a way to turn convergence in $L^2$ topology
into pointwise convergence in $t$ by integrating against
another $L^2$ function.  In mathematics, there is a
rigorous real-time, time-sliced path integral where
the convergence of the improper Lebesgue integrals are
taken in the $L^2$ topology.  We will use the above idea
to convert the $L^2$ convergence path integral into
improper Riemann integrals.

\subhead
{\bf 2. Background}\endsubhead
In this section, we will provide some background
with references; we leave it to the reader to
look up the proofs.
The following is a generalized Trotter
product formula due to Chernoff([8]).  We will
use a modified version of the theorem to prove (1.5).  
\proclaim{\bf Theorem 2.1} Let $F\left(t\right), t\geq 0$ be a family of
linear contractions in $X$ with $F\left(0\right) = I$.  If the closure
$C$ of the strong derivative
$$
  F^{'}\left(0\right) = s-\lim_{\epsilon\to 0}
  \dfrac{F\left(\epsilon\right) - I}{\epsilon}\tag2.1
$$
generates a contractive semigroup, then 
$F\left(\frac{t}{k}\right)^k \to e^{tC}$
strongly and uniformly on bounded $t$ intervals.
\endproclaim
\demo{Proof}
See [8].\qed
\enddemo

We are interested in the essential self-adjointness
of (1.3); the following two theorems provide
conditions on $\vec a$ and $V$ for essential
self-adjointness.  For more details on
the subject, see [9], [12], [14], [19] and references within.  
\proclaim{\bf Theorem 2.2} Let $n \geq 4$ and $p = \frac{6n}{n+2}$. 
Let $\vec a$ be an $n$ dimensional vector in $\Bbb R^n$ 
and 
each component of $\vec a$ is in $L_{loc}^p$.  Furthermore,
let $V, \nabla\vec a\in L_{loc}^{\frac{p}{2}}$, then 
$\left(-i\nabla - \vec a\right)^2 + V$ is essentially self-adjoint
on $C_0^\infty\left(\Bbb R^n\right)$
\endproclaim
\demo{Proof} See [19].\qed
\enddemo

\proclaim{\bf Theorem 2.3} Suppose that each 
component of $\vec a$ is a real-valued
functions in $L^4\left(\Bbb R^3\right) + L^\infty\left(\Bbb R^3\right)$,
$\nabla * \vec a\in L^2\left(\Bbb R^3\right) + 
L^\infty\left(\Bbb R^3\right)$,
and $V$ is a real-valued function in $L^2 + L^\infty$.  For 
$\phi\in C_0^\infty\left(\Bbb R^3\right)$, define
$$
  H\phi = -\Delta\phi + 2i\vec a * \nabla\phi + i  
  \left(\nabla * \vec a\right)\phi + V\phi + \vec a^2\phi, \tag2.2
$$
then $H$ is essentially self-adjoint on 
$C_0^\infty\left(\Bbb R^3\right)$.
\endproclaim
\demo{Proof} See [17].\qed
\enddemo

The following lemma is due to Simon([19]).  We will use the
idea to show (1.5). 
\proclaim{\bf Lemma 2.4} Let $a_j\in L_{loc}^2\left(\Bbb R^n\right)$.
Then $-i\partial_j - a_j$ is essentially self-adjoint
on $C_0^\infty\left(\Bbb R^n\right)$ and its closure
$-iD_j$ obeys
$$
  -iD_j = e^{i\lambda_j}\left(-i\partial_j\right)e^{-i\lambda_j}\tag2.3
$$
for a real-valued function $\lambda_j \in 
L_{loc}^2\left(\Bbb R^n\right)$.  Furthermore, the
domain of $D_j$ is
$$
  \Cal D_j\left(D_j\right) =
  \left\{\phi\in L^2|\left(\partial_j - ia_j\right)\phi
  \left(\text{dist. sense}\right)\in L^2\right\}\tag2.4
$$
and
$$
  \lambda_j\left(x_1,\dots ,x_n\right) =
  \int_0^{x_j} a_j\left(x_1,\dots ,x_{j-1}, y,\dots ,x_n\right)dy .\tag2.5
$$
\endproclaim  
\demo{Proof} 
See [19]. \qed 
\enddemo
Since $L_{loc}^p\left(\Bbb R^n\right) \subset
L_{loc}^q\left(\Bbb R^n\right)$ for $p \geq q$,
any $\vec a$ that satisfies theorem 2.2 or theorem
2.3 will satisfy lemma 2.4.  
For any $\vec a$ that satisfies theorem 2.2 or theorem
2.3, lemma 2.4 implies that 
for $\phi\in C_0^\infty\left(\Bbb R^n\right)$, the 
following holds
$$\align
  {}&\left[e^{i\lambda_j}
  \left(-i\partial_j\right)e^{-i\lambda_j}\right]^2\phi = 
  e^{i\lambda_j}\left(-\partial_j^2\right)e^{-i\lambda_j}\phi =
  \left(-i\nabla_j - a_j\right)^2\phi = \tag2.6 \\
  {}&-\Delta{}_{j}\phi + 2ia_j\nabla_j\phi + i
    \left(\nabla_j a_j\right)\phi + a_j^2\phi  
    \in L^2\left(\Bbb R^n\right).
\endalign
$$ 
Let us denote $H_0^j\phi = -\Delta_j\phi$ for $\phi\in
C_0^\infty\left(\Bbb R^n\right)$, then we can write
$$
  -ie^{i\lambda_j}\left(-\partial_j^2\right)e^{-i\lambda_j}\phi = 
  \text{l.i.m}_{\epsilon\to 0}\left\{
  e^{i\lambda_j}\dfrac{\text{exp}\left(
  -i\epsilon H_0^j\right) - I}{\epsilon}
  e^{-i\lambda_j}\phi\right\}. \tag2.7
$$

\proclaim{\bf Proposition 2.5} Let $\vec a$ and $V$ be as in theorem 2.2
or theorem 2.3.
Denote 
$$
  F_j\left(t\right) = e^{-itV}\prod_{k = 1}^j
  e^{i\lambda_k}\text{exp}
  \left(-it H_0^k\right)e^{-i\lambda_k}, \tag2.8
$$
then for $\phi\in C_0^\infty\left(\Bbb R^n\right)$,
$$
  F_j^{'}\left(0\right)\phi = -i\left(\sum_{k = 1}^j
  \left(-i\nabla_k - a_k\right)^2 + V\right)\phi ,\,  
  1 \leq j\leq n\tag2.9
$$
\endproclaim
\demo{Proof} Equation (2.6) and (2.7) implies that (2.9) is
true for $j = 1$.  We will show (2.9) by induction up to $j = n$.
Let us denote 
$$\align
  {}&\bar F\left(\epsilon\right) = e^{-i\epsilon V}
  \text{exp}\left(i\lambda_{j+1}\right)\text{exp}
  \left(-i\epsilon H_0^{j+1}\right)
  \text{exp}\left(-i\lambda_{j + 1}\right),\tag2.10 \\
  {}&\bar F_j\left(\epsilon\right) = 
    \prod_{k = 1}^j
    e^{i\lambda_k}\text{exp}
    \left(-i\epsilon H_0^k\right)e^{-i\lambda_k}. 
\endalign
$$
Suppose (2.9) is true for $1 \leq j < n$, then
$$\align
  {}&F_{j+1}^{'}\left(0\right)\phi = \text{l.i.m}_{\epsilon\to 0}
     \dfrac{\left(\bar F\left(\epsilon\right)
     \bar F_j\left(\epsilon\right)- I
     \right)\phi}{\epsilon} = \tag2.11\\  
  {}&\text{l.i.m}_{\epsilon\to 0}
     \dfrac{\left(\bar F\left(\epsilon\right)
     \bar F_j\left(\epsilon\right) - \bar F\left(\epsilon\right)
     \right)\phi}{\epsilon}\, + 
    \text{l.i.m}_{\epsilon\to 0}
    \dfrac{\left(\bar F\left(\epsilon\right)
     - I\right)\phi}{\epsilon} = \\
  {}&-i\left(\sum_{k = 1}^{j + 1}
     \left(-i\nabla_k - a_k\right)^2 + V\right)\phi .\qed
\endalign
$$
\enddemo
We can not 
immediately apply theorem 2.1 on this $F_n\left(t\right)$ 
to produce a Trotter product formula for (1.5) since we
only know the behavior of $F^{'}\left(0\right)$
on $C_0^\infty$ while theorem 2.1 requires
knowledge of the behavior of $F^{'}\left(0\right)$
on its domain.  We will prove a slightly different 
version of theorem 2.1 which only requires 
knowledge of $F^{'}\left(0\right)$ on
a dense subset of its domain.  The proof is 
a small modification of Chernoff's proof
of theorem 2.1

\subhead
{\bf 3. Generalized Trotter Product formula}\endsubhead
We quote a few results from Chernoff([8]) ; we leave it to
the reader to look up the proofs.  
\proclaim{\bf Lemma 3.1} Let $C_n, n = 1, 2,\dots$, and
$C$ be the generators of $\left(C_0\right)$ contraction
semigroups on $X$.  Let $\Cal D$ be a dense subspace of the
domain of $C$ such that $\overline {C|\Cal D} = C$.  Suppose
that for all $\phi\in\Cal D, C_n\phi$ is defined and
$\lim_{n\to\infty}C_n\phi = C\phi$.  Then for every
$\lambda > 0$, $\left(\lambda - C_n\right)^{-1}$
converges to $\left(\lambda - C\right)^{-1}$ in the 
strong operator topology.  
\endproclaim 
\demo{Proof} See [8].\qed
\enddemo

\proclaim{\bf Proposition 3.2} Under the hypothesis of lemma
3.1, $e^{tC_n}\to e^{tC}$ in the strong operator topology,
the convergence being uniform on every compact interval.
\endproclaim
\demo{Proof} See [8].\qed
\enddemo

\proclaim{\bf Lemma 3.3} Let $T$ be a linear contraction on $X$.
Then $t\to e^{t\left(T - I\right)}$ is a contraction
semigroup.  For all $\phi\in X$ we have 
$||\left(e^{t\left(T - I\right)} - T^k\right)\phi||
\leq k^{\frac{1}{2}}||\left(T - I\right)\phi||$.
\endproclaim
\demo{Proof} See [8].\qed
\enddemo

The following is the result that we seek. 
\proclaim{\bf Theorem 3.4} Let $F\left(t\right)$ be
a strongly continuous function from $\left[0, \infty\right)$
to the linear contractions on $X$ such that $F\left(0\right)
= I$.  Suppose that $F^{'}\left(0\right)$ is defined
on a dense subset $\Cal D$ of $X$ and its closure $C$
generates a contractive semigroup.  Then 
$F^k\left(\frac{t}{k}\right)$ converges to $e^{tC}$ in
the strong operator topology.  
\endproclaim
\demo{Proof} Fix $t > 0$.  Define $C_k = 
   \dfrac{k\left(F\left(\frac{t}{k}\right) - I\right)}{t}$.
Lemma 3.3 implies that 
$$
  e^{k\left(F\left(\frac{t}{k}\right) -
  I\right)} = e^{tC_k} \tag3.1
$$ 
exists.  The hypotheses of lemma 3.1
are satisfied, hence proposition 3.2 implies that for all
$\phi\in X$,
$$
  e^{tC_k}\phi \to e^{tC}\phi \tag3.2
$$
Suppose that $\phi\in\Cal D$.  Lemma 3.3 implies that
$$\align
  {}&\bigg|\bigg|e^{tC_k}\phi - F^k
     \left(\frac{t}{k}\right)\phi \bigg|\bigg|
     \leq k^{\frac{1}{2}}
     \bigg|\bigg|\left(F\left(\frac{t}{k}\right) - 
     I\right)\phi \bigg|\bigg| = \tag3.3 \\
  {}&\dfrac{t}{k^{\frac{1}{2}}}\bigg|\bigg|
     \dfrac{k}{t}\left(
      F\left(\frac{t}{k}\right) - I
      \right)\phi\bigg|\bigg|\to 0 \text{ as } k\to\infty. 
\endalign
$$
Hence for $\phi\in\Cal D$, $F^k\left(\frac{t}{k}\right)\phi
\to e^{tC}\phi$.  Since the operators are contractions,
we have the result of all $\phi\in X$.  \qed
\enddemo
The proof of theorem 3.4 is exactly that of theorem
3.1 except that Chernoff let $\Cal D$ be the 
domain of $F^{'}\left(0\right)$ where as we took
$\Cal D$ to be any dense subset of $X$ for which
$F^{'}\left(0\right)$ is well defined.  We can
apply theorem 3.4 to the operator $F_n$ as defined in
proposition 2.5.  

\proclaim{\bf Theorem 3.5}  Let $\vec a$ and $V$ satisfy
the conditions of either theorem 2.2 or theorem 2.3.  Let
$F\left(t\right) =
F_n\left(t\right)$ where $F_n\left(t\right)$ is 
as defined in proposition 2.5 and denote 
$\bar H$ to be the closure of $H$, then for all $\phi\in
L^2\left(\Bbb R^n\right)$,
$$
  \text{l.i.m}_{k\to\infty}F^k\left(\dfrac{t}{k}\right) 
  \phi = e^{-it\bar H}\phi .\tag3.4 
$$
\endproclaim
\demo{Proof} Proposition 2.5 implies that $F\left(t\right)$
satisfies the conditions in theorem 3.4 with
$
  F^{'}\left(0\right) = -iH
$
and $\Cal D$ = $C_0^\infty\left(\Bbb R^n\right)$.  The
closure of $
  F^{'}\left(0\right) = -iH
$ is $-i\bar H$ and it generates a contractive semigroup
$e^{-it\bar H}$.  Hence 3.4 follows.  \qed
\enddemo 

\subhead
{\bf 4. Rigorous real-time 
Feynman path integral for evolution}\endsubhead 
We are now ready to derive Feynman path integrals for vector potential
Hamiltonians.  
We will assume that
$\vec a$ and $V$ have at most a finite number of discontinuities
and singularities.  Suppose we take two $L^2$ functions 
$\phi$ and $\psi$ such that they have at most a finite number
of singularities and discontinuities and that
$\left\{\vec w_1,\dots ,\vec w_p\right\}$ are all the
singular and discontinuous points of $\vec a, V, \phi,$
and $\psi$.  For $1\leq \alpha\leq p$ and $1\leq \beta\leq n$,
let $w_\alpha^\beta$ be the $\beta$th coordinate of $\vec w_\alpha$.
For any fixed $k\in \Bbb N\bigcup\left\{0\right\},$
let $a_\beta^k ,a_\beta^{k}, b_\beta^{\alpha, k},$ and  
$d_\beta^{\alpha, k}
\comment
a_{\beta ,m}^k ,\bar a_{\beta ,m}^{k}, b_{\beta ,m}^{\alpha, k},
d_{\beta ,m}^{\alpha, k},
a_{\beta ,n}^k ,a_{\beta ,n}^{k}, b_{\beta ,n}^{\alpha, k},$
and $d_{\beta ,n}^{\alpha, k}, 
\endcomment 
\in\Bbb R^{+}$.  
Let 
$$C_{\left\{j_\beta^k\right\}} = 
\left(-a_\beta^k, \bar a_\beta^k\right) - 
\bigcup_{\alpha = 1}^p \left(w_\alpha^\beta - 
\frac{1}{b_\beta^{\alpha , k}},
w_\alpha^\beta + 
\frac{1}{d_\beta^{\alpha , k}}\right). \tag4.1$$  
We will use the notations,  
$$
  \left\{j_\beta^k\right\} = 
   \left\{a_\beta^k ,\bar a_\beta^k, b_\beta^{1,k},
    d_\beta^{1,k},\dots , b_\beta^{p,k}, d_\beta^{p,k}\right\},
   \tag4.2
$$
and for any set of numbers $\left\{J\right\}$, we will
denote by $\left\{J\right\}\to\infty$ to mean that
each element of the set goes to infinity 
independent of each other.  
Notice that if we let 
$\left\{j_\beta^k\right\}\to\infty$,
we obtain $\Bbb R - \left\{\vec w_1,\dots ,\vec w_p\right\}$ 
in (4.1).
Furthermore,
let
$$\align
  {}&C_{\left\{j_k\right\}} = C_{\left\{j_1^k\right\}}\times
  \dots\times C_{\left\{j_n^k\right\}}, \tag4.3 \\
  {}&\left\{j_k\right\} = \bigcup_{\beta = 1}^n
     \left\{j_\beta^k\right\}, 
\endalign
$$
and
$$
  \align
  {}&D_k = C_{\left\{j_0\right\}}\times\dots\times
     C_{\left\{j_k\right\}}, \tag4.4 \\
  {}&\left\{J_k\right\} = \bigcup_{l = 0}^k
     \left\{j_l\right\}.  
\endalign
$$
We will denote the characteristic function of
$C_{\left\{j_l^k\right\}}, C_{\left\{j_k\right\}},$
and $D_k$
by $\chi_{C_{\left\{j_l^k\right\}}},$ 
$\chi_{C_{\left\{j_k\right\}}},$ and 
$\chi_{D_k}$ 
respectively.  With the above notation, for
$\psi\in L^2$, we can write
$$\align
  {}&F\left(\frac{t}{k}\right)\psi = 
  \text{exp}\left\{\dfrac{-itV}{k}\right\}\prod_{l = 1}^n
  e^{i\lambda_l}\text{exp}
  \left(\dfrac{-it H_0^l}{k}\right)e^{-i\lambda_l}\psi = \tag4.5 \\
  {}&\text{exp}\left\{\dfrac{-itV}{k}\right\}
  \prod_{l = 1}^n
  e^{i\lambda_l}\text{exp}
  \left(\dfrac{-it H_0^l}{k}\right)
  \text{l.i.m}_{\left\{j_l^{0}\right\}\to\infty}
  \chi_{C_{\left\{j_l^{0}\right\}}}
  e^{-i\lambda_l}\psi = \\
  {}&\text{l.i.m}_{\left\{j_{0}\right\}\to\infty}
  \text{exp}\left\{\dfrac{-itV}{k}\right\}
  \prod_{l = 1}^n
  e^{i\lambda_l}\text{exp}
  \left(\dfrac{-it H_0^l}{k}\right)
  \chi_{C_{\left\{j_l^{0}\right\}}}
  e^{-i\lambda_l}\psi .
\endalign
$$
In (4.5), all limits are taken independent of each other 
and the second equality is due to the fact that 
all operators are continuous from $L^2$ to $L^2$.  
For $1\leq l \leq n,
\epsilon = \frac{t}{k},$ and 
$\phi\in L^2$, we can
write 
$$\align
  {}&e^{i\lambda_l}\text{exp}
  \left(-i\epsilon H_0^l\right)
  \chi_{C_{\left\{j_l^{0}\right\}}}
  e^{-i\lambda_l}\phi = \tag4.6\\
  {}&\left(\dfrac{1}{4i\pi\epsilon}\right)^
    \frac{1}{2}\text{exp}\left[i
    \int_0^{x_1^l} a_l\left(x_{0}^1,\dots ,x_{0}^{l-1}, 
     y,\dots ,x_{0}^n\right)dy\right]* \\
  {}&\int\limits_{C_{\left\{j_l^{0}\right\}}}
     \Bigg\{\text{exp}\left[\frac{i\epsilon}{4}
     \left(\frac{x_1^{l} - x_{0}^l}{\epsilon}\right)^2\right]\\
  {}&
     \text{exp}\left[-i
     \int_0^{x_{0}^l} a_l\left(x_{0}^1,\dots ,x_{0}^{l-1},
     y,\dots ,x_{0}^n\right)dy\right]
     \phi\left(\vec x_{0}\right)\Bigg\}dx_{0}^l = \\
  {}&\left(\dfrac{1}{4i\pi\epsilon}\right)^
    \frac{1}{2}\int\limits_{C_{\left\{j_l^{0}\right\}}}
    \text{exp}\left[\frac{i\epsilon}{4}
     \left(\frac{x_1^{l} - x_{0}^l}{\epsilon}\right)^2\right]
     \text{exp}\left[i\bar \lambda_l
     \left(\vec x_1, \vec x_{0}\right)\right]
     \phi\left(\vec x_{0}\right)dx_{0}^l , 
\endalign
$$
where $x_{0}^j, x_{1}^j$ is the $jth$ coordinate of the vector
$\vec x_{0}$ and $\vec x_{1}$ respectively and 
$$\align
   {}&\bar \lambda_l
     \left(\vec x_1, \vec x_{0}\right) = \tag4.7\\
   {}&\int_0^{x_1^l} a_l\left(x_{0}^1,\dots ,x_{0}^{l-1},
     y,\dots ,x_{0}^n\right)dy - 
     \int_0^{x_{0}^l} a_l\left(x_{0}^1,\dots ,x_{0}^{l-1},
     y,\dots ,x_{0}^n\right)dy = \\ 
   {}&\int_{x_{0}^l}^{x_1^l} a_l\left(x_{0}^1,\dots ,x_{0}^{l-1},
      y,\dots ,x_{0}^n\right)dy
\endalign
$$ 
Equations (4.5) and (4.6) implies that
$$\align
  {}&F\left(\frac{t}{k}\right)\psi = 
    \text{l.i.m}_{\left\{j_{0}\right\}\to\infty}\Bigg\{
    \text{exp}\left\{\dfrac{-itV\left(\vec x_1\right)}{k}\right\}
    \left(\dfrac{1}{4i\pi\epsilon}\right)^
    \frac{n}{2}\tag4.8\\
  {}&\int\limits_{C_{\left\{j_0\right\}}}
    \text{exp}\left[\frac{i\epsilon}{4}
     \left(\frac{\vec x_1 - \vec x_{0}}{\epsilon}\right)^2 + 
     i\sum_{l = 1}^n\bar \lambda_l
     \left(\vec x_1, \vec x_{0}\right)\right]
     \psi\left(\vec x_0\right)d\vec x_0\Bigg\} = \\
  {}&\text{l.i.m}_{\left\{j_{0}\right\}\to\infty}\Bigg\{
    \text{exp}\left\{\dfrac{-itV\left(\vec x_1\right)}{k}\right\}
    \left(\dfrac{1}{4i\pi\epsilon}\right)^
    \frac{n}{2}\\
  {}&\int\limits_{C_{\left\{j_0\right\}}}
    \text{exp}\left[\frac{i\epsilon}{4}
     \left(\frac{\vec x_1 - \vec x_{0}}{\epsilon}\right)^2 +
    i\bar \lambda\left(\vec x_1, \vec x_{0}\right)\right]
     \psi\left(\vec x_0\right)d\vec x_0\Bigg\}, 
\endalign
$$
where $\bar \lambda\left(\vec x_1, \vec x_{0}\right) =
\sum_{l = 1}^n\bar \lambda_l
     \left(\vec x_1, \vec x_{0}\right)$, and
all limits are taken independent of each other.  
As mentioned earlier, if the vector potential is well
behaved and if $\vec x_1, \vec x_{0}$ are close, we
can approximate the last integral in 4.7 and conclude that
$\bar \lambda\left(\vec x_{1}, \vec x_0\right)$
is close to $\left(\vec x_{1} - \vec x_0\right)*\vec a\left(
     \dfrac{\vec x_{1} + \vec x_0}{2}\right)$.

here i am Furthermore, 
$$\align
    {}&F^k\left(\frac{t}{k}\right)\psi = \tag4.9 \\ 
    {}&
    \text{l.i.m}_{\left\{J_{k-1}\right\}\to\infty}
    \left(\dfrac{1}{4i\pi\epsilon}\right)^
    \frac{n\left(k - 1\right)}{2}
    \int\limits_{D_{k - 1}} 
     \text{exp}\Bigg\{i\epsilon
      \sum_{j = 0}^{k-1}\bigg[\dfrac{1}{4}
     \left(\dfrac{\vec x_{j+1} - \vec x_j}{\epsilon}\right)^2 
     - \\
  {}&V\left(\vec x_{j+1}\right) + 
     \dfrac{\bar \lambda\left(\vec x_{j+1}, \vec x_j\right)}
     {\epsilon}
     \bigg]\Bigg\}\psi\left(\vec x_0\right)
     d\vec x_0\dots d\vec x_{k-1} = 
     \left(\dfrac{1}{4i\pi\epsilon}\right)^
    \frac{n\left(k - 1\right)}{2}\times \\
  {}&\text{l.i.m}_{\left\{J_{k-1}\right\}\to\infty}
    \int\limits_{D_{k - 1}}
     \text{exp}\left\{i\epsilon
     S_k\left(\epsilon ,\vec x_0\dots \vec x_k\right)\right\}
     \psi\left(\vec x_0\right)
     d\vec x_0\dots d\vec x_{k-1},
\endalign
$$
where all limits are again taken independent of each other.  
We see that the vector potential term in (4.9) is nowhere
similar to that of (1.1).  

Let us use the notation $\int_{rK}$ to denote Riemann or improper
Riemann integral over $K$ and $\int_{K}$ 
to denote Lebesgue integral over $K$.  We now prove our 
main result in (1.2).  

\proclaim{\bf Theorem 4.1} Let $\phi ,\psi\in L^2$ be such
that they each have at most a finite number of discontinuities
and singularities.  Suppose $\vec a$ and $V$ satisfies
theorem 2.2 or theorem 2.3 and that $\vec a$ and $V$
has at most a finite number of discontinuities and singularities.
With our previously defined notations,
the following
holds
$$\align
   {}&\Bigg<\phi^{*},
      \text{exp}\left(-it\bar H\right)
       \psi\Bigg>_{L^2} =
    \int_{\Bbb R^n} \phi\left(\vec x\right)\left[
    \text{exp}\left(-it\bar H\right)
    \psi\right]\left(\vec x\right)
    d\vec x = \tag4.10\\
   {}&\lim_{k\to\infty}\Bigg\{\left(\dfrac{1}{4i\pi\epsilon}\right)^
    \frac{n\left(k - 1\right)}{2}
    \int\limits_{r\Bbb R^{n(k + 1)}}\phi\left(\vec x_k\right)
    \times\\
   {}&\text{exp}\Bigg\{i\epsilon
    \sum_{j = 0}^{k - 1}\left[\dfrac{1}{4}
     \left(\dfrac{\vec x_{j+1} - \vec x_j}{\epsilon}\right)^2 -
     V\left(\vec x_{j+1}\right) +
     \dfrac{\bar \lambda\left(\vec x_{j+1}, \vec x_j\right)}
     {\epsilon}
     \right]\Bigg\}\psi\left(\vec x_0\right)
     d\vec x_0\dots d\vec x_{k}\Bigg\}.
\endalign
$$
\endproclaim
\demo{Proof} Theorem 3.5 and Schwarz's inequality
imply that 
$$\align
   {}&\Bigg<\phi{}^* ,
    \text{exp}\left(\dfrac{-it\bar H}{\hbar}\right)
     \psi\Bigg>_{L^2} = \Bigg<\phi{}^* ,
     \text{l.i.m}_{k\to\infty}
     F^k\left(\dfrac{t}{k}\right)\psi\Bigg>_{L^2} = \tag4.11\\
   {}&\lim_{k\to\infty}
      \Bigg<\phi{}^* ,
       F^k\left(\dfrac{t}{k}\right)\psi\Bigg>_{L^2}.  
\endalign
$$      
For any $k\in\Bbb N$ with $\epsilon = \frac{t}{k}$, 
(4.9) implies that 
$$\align
  {}&\Bigg<\phi{}^* ,
       F^k\left(\dfrac{t}{k}\right)\psi\Bigg>_{L^2} =
    \left(\dfrac{1}{4i\pi\epsilon}\right)^
    \frac{n\left(k - 1\right)}{2} 
    \int\limits_{\Bbb R^n}\left[ 
     \text{l.i.m}_{\left\{j_k\right\}
     \to\infty}\chi_{C_{\left\{j_k\right\}}}
     \phi\left(\vec x_k\right)\right]\times \tag4.12\\
  {}&\bigg\{\text{l.i.m}_{\left\{J_{k-1}\right\}\to\infty}
    \int\limits_{D_{k - 1}}
     \text{exp}\left\{i\epsilon
     S_k\left(\epsilon ,\vec x_0\dots \vec x_k\right)\right\}
     \psi\left(\vec x_0\right)
     d\vec x_0\dots d\vec x_{k-1}\bigg\}d\vec x_k = \\
  {}&\left(\dfrac{1}{4i\pi\epsilon}\right)^
    \frac{n\left(k - 1\right)}{2}
    \lim_{\left\{J_{k}\right\}\to\infty}
    \int\limits_{D_{k}}\phi\left(\vec x_k\right)
    \text{exp}\left\{i\epsilon
     S_k\left(\epsilon ,\vec x_0\dots \vec x_k\right)\right\}
     \psi\left(\vec x_0\right)
     d\vec x_0\dots d\vec x_{k} , 
\endalign
$$       
in the last equality of (4.12), we used Schwarz's inequality
and took all the $L^2$
limits outside of the integral as pointwise limits on
$t$.  Notice that all limits in (4.12) 
are taken independent of each other.  By construction of 
$D_{k}$ and the hypothesis on $\vec a, V, \phi,$ and $\psi$,
we see that $\phi\left(\vec x_k\right)\text{exp}\left\{i\epsilon
S_k\left(\epsilon ,\vec x_0\dots \vec x_k\right)\right\} 
\psi\left(\vec x_0\right)$ is a bounded and continuous function
on $D_{k}$.  Thus, we can replace the Lebesgue integral
over $D_{k}$ by a Riemann integral over $D_{k}$.  By construction
of $D_{k}$ and the fact that all limits are taken independent
of each other, we can interpret the limits as improper
Riemann integrals.  Hence, for all $k\in\Bbb N$,
$$\align
  {}&\Bigg<\phi{}^* ,
       F^k\left(\dfrac{t}{k}\right)\psi\Bigg>_{L^2} = \tag4.13\\
  {}&\left(\dfrac{1}{4i\pi\epsilon}\right)^
    \frac{n\left(k - 1\right)}{2}
    \int\limits_{r\Bbb R^{nk}}
    \phi\left(\vec x_k\right)
    \text{exp}\left\{i\epsilon
     S_k\left(\epsilon ,\vec x_0\dots \vec x_k\right)\right\}
     \psi\left(\vec x_0\right)
     d\vec x_0\dots d\vec x_{k}.
\endalign
$$
The theorem follows from equations (4.13) and (4.11). \qed
\enddemo

\enddocument